# Improved Watermarking Scheme Using Discrete Cosine Transform and Schur Decomposition

[1] Henri Bruno Razafindradina , [2] Nicolas Raft Razafindrakoto , [3] Paul Auguste Randriamitantsoa

[1] Higher Insitute of Technology, Diégo-Suarez, 201, Madagascar

[2,3] Polytechnic High School of Antananarivo, University of Antananarivo, Antananarivo, 101, Madagascar

**Abstract**
Watermarking is a technique which consists in introducing a brand, the name or the logo of the author, in an image in order to protect it against illegal copy. The capacity of the existing watermark channel is often limited. We propose in this paper a new robust method which consists in adding the triangular matrix of the mark obtained after the Schur decomposition to the DCT transform of the host image. The unitary matrix acts as secret key for the extraction of the mark. Unlike most watermarking algorithms, the host image and the mark have the same size. The results show that our method is robust against attack techniques as : JPEG compression, colors reducing, adding noise, filtering, cropping, low rotations, and histogram spreading.

*Keywords:* *DCT, copyright, watermarking, Schur.*

## 1. Introduction

Since the standardization of JPEG, the number of digital media circulating on the Internet continues to increase. Nowadays, it's very easy to obtain the tools for copying, manipulating these medias. It became extremely easy to reproduce any medium. In the case of digital media (sound, image and video), the research is directed towards a technical resolution by inserting a mark in the medium in the order to identify its owner. These techniques must observe four conditions : the imperceptibility of the mark (the mark does not affect the image quality and the image itself still has its commercial quality after being watermarked), the robustness (watermark must resist any voluntary or involuntary attacks), the security (insertion and extraction must use a secret key to demotivate attacks) and the capacity. The last criterion is generally overlooked ; it defines the number of bits of the mark that can be inserted into the host image.

Watermarking in the DCT domain was initiated by Cox and al [1][2], Koch and Zhao [3][4] were followed by many researchers to optimize a compromise imperceptibility / robustness. These algorithms had a low capacity like the Koch and Zhao scheme. Variants of watermarking in DCT domain have been proposed :

Hsu [5] swapped the mark coefficients in order to make the technique robust against crop. Fotopoulos [6] combine Wavelets and DCT transform : it inserts the mark in the DCT coefficients of each sub-band after Wavelets decomposition. Saryazdi and Al [7] have tried to improve the Cox additive scheme based on the AC estimated coefficients from DC. Some like Golikeri and Al [8] followed the same technique by inserting the mark in the DCT blocks with high energy. Despite all these efforts, the capacity of watermarking remained very limited. To improve this criterion, Sharkas and Al proposed [9] a technique which consists in inserting a mark in another mark before being inserted in the host image. Many recent watermarking techniques are inspired by the usual compression method and matrix decomposition. These includes blind SVD technique [10] that inserts the bits of the mark in the singular values matrix, A combination of DCT and SVD [11], DCT and DWT [12][13], spatial and frequency domain [14], improving the Al-Haj [15], DCT-DWT-SVD [16], DCT-VQ [17] methods aiming to make watermarking techniques robust against most attacks. In addition, the number of publications based on the SVD techniques shows a great interest of matrix decomposition in watermarking.

In this paper, we present a new approach based on DCT and the Schur decomposition. The basic principles of Schur decomposition and DCT transform are first recalled, the proposed method is detailed and finally, the experimental results will be discussed.

## 2. Definition

### 2.1 Schur decomposition

Schur's theorem announced that : if $A \in C^{n \times n}$, there exists a unitary matrix U and an upper triangular matrix T, such that :

$$A = U \times T \times U' \qquad (1)$$

where $T=\Lambda+N$ with $\Lambda=\text{diag}(\lambda_1,\ldots,\lambda_n)$ diagonal matrix of eigenvalues of A and N an upper triangular matrix.

## 2.2 Specifications

Unlike SVD decomposition, Schur does not require very high computing power.

## 2.3 DCT transform

The DCT of an image which is a variant of the Fourier transform is used to describe each n×n block in a frequencies map and amplitudes rather than pixels and colors. The value of a frequency reflects the importance of a rapid change, while the value of magnitude corresponds to the deviation associated with each color change.

For each n×n pixels block are thus associated n×n frequencies.

The DCT is expressed mathematically by :

$$DCT(i,j) = \frac{2}{n} C(i)C(j) \sum_{x=0}^{n-1}\sum_{y=0}^{n-1} p(x,y) \cos\left[\frac{(2x+1)i\pi}{2n}\right]\cos\left[\frac{(2y+1)j\pi}{2n}\right] \quad (2)$$

where $p(x,y)$ : pixel of an image
And inverse DCT transform is expressed by :

$$p(x,y) = \frac{2}{n} \sum_{i=0}^{n-1}\sum_{j=0}^{n-1} C(i)C(j) DCT(i,j) \cos\left[\frac{(2x+1)i\pi}{2n}\right]\cos\left[\frac{(2y+1)j\pi}{2n}\right] \quad (3)$$

In both cases, the constant C is defined by :

$$C(x) = \begin{cases} \frac{1}{\sqrt{2}} & \text{when } x = 0 \\ 1 & \text{when } x > 0 \end{cases}$$

## 2.4 Specifications

Generally, most of the energy is concentrated in the low frequencies of the spectrum which are expressed by the closest coefficients to the upper left corner of the transformed matrix.

## 3. Proposed Watermarking Scheme

### 3.1 Watermark insertion

Let I the host image and W the mark to insert. The two images are the same size. The first stage of the technique is to compute the DCT transform noted $I_{DCT}=DCT(I)$ of the host image. Then, the Schur decomposition is applied to the mark and we obtain $U_W$ and $T_W$, such : $W=U_W \times T_W \times U_W'$. $U_W$ matrix is kept as a private key and we add $T_W$ to the $I_{DCT}$ matrix according to the formula :

$$I_{DCTW} = I_{DCT} + \alpha \times T_W \quad (4)$$

The choice of the $\alpha$ coefficient depends on the compromise strength / imperceptibility desired by the user.

Finally, the watermarked image is obtained by computing the inverse transform of $I_{DCTW}$, we write :

$$I_W = DCT^{-1}(I_{DCTW}) \quad (5)$$

The insertion and detection algorithm are described in the following figure :

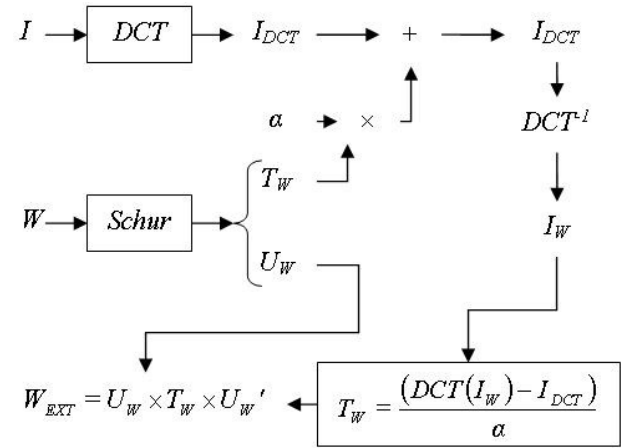

Fig. 1  Insertion and Detection

### 3.1 Watermark detection

Detection consists to compute $T_W$ and rebuild the mark with formulas 6 and 7 :

$$T_W = \frac{(DCT(I_W) - I_{DCT})}{\alpha} \quad (6)$$

$$W_{EXT} = U_W \times T_W \times U_W' \quad (7)$$

where $W_{EXT}$ is the extracted watermark

## 4. Results and Discussion

To test our technique, we used images of size 512×512 (see Table 1). In order to ensure a best robustness / imperceptibility ratio, we have chosen $\alpha=0,3$, except for the first $\lambda_1$ coefficient which is equal to 0,03. This choice ensures imperceptibility of the mark.

### 4.1 Imperceptibility: Evaluating distortion

The PSNR (Peak Signal to Noise Ratio) is traditionally used to express the distortion or the impact of the mark insertion. It's expressed by the following formula :

$$PSNR = 10 \times \log_{10}\left(\frac{Max[I(x,y)]^2}{MSE}\right) \quad (8)$$

where : MSE is the Mean Square Error

$$MSE = \frac{1}{n \times n} \sum_{x=1}^{n} \sum_{y=1}^{n} [I(x,y) - I_W(x,y)]^2 \qquad (9)$$

The following figures represent respectively the host image, the inserted mark, the watermarked image and the difference between original and watermarked image.

The obtained PSNR is 30 dB, which corresponds to an insertion strength equal to "30" [18].

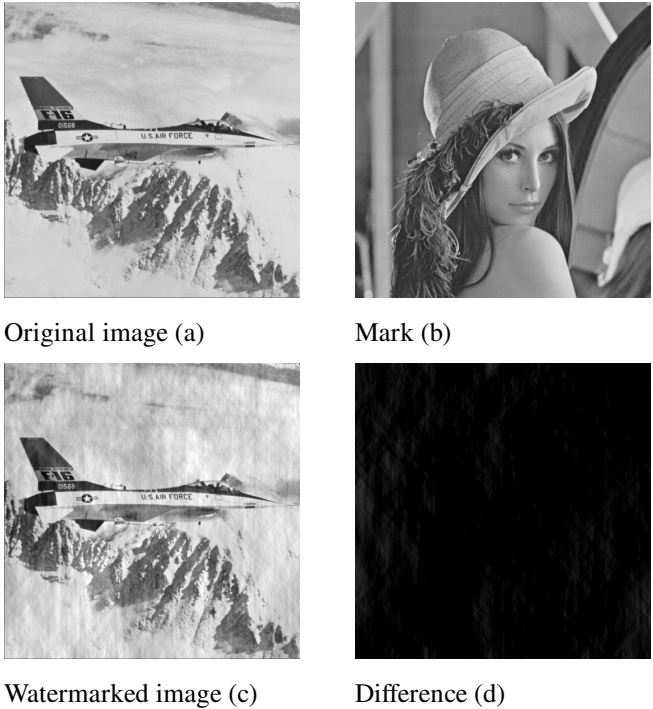

Original image (a)        Mark (b)

Watermarked image (c)        Difference (d)

Fig. 2  a and b) Host image and mark, c and d) Watermarked image and the difference between 2 images

### 4.2 Detection and Robustness

Ideally, given a watermarked image, an unauthorized entity should not be able to destroy the mark (the mark must resist common signal processing and intentional attacks).

The robustness of the method is evaluated by the correlation between the original W and the extracted mark $W_{EXT}$ by computing :

$$corr(W, W_{EXT}) = \frac{\sum_{x=1}^{n}\sum_{y=1}^{n}[W(x,y)-\overline{W}][W_{EXT}(x,y)-\overline{W}_{EXT}]}{\sqrt{\left(\sum_{x=1}^{n}\sum_{y=1}^{n}[W(x,y)-\overline{W}]^2\right)\left(\sum_{x=1}^{n}\sum_{y=1}^{n}[W_{EXT}(x,y)-\overline{W}_{EXT}]^2\right)}} \qquad (10)$$

where W and $W_{EXT}$ are respectively the means of mark W and extracted mark $W_{EXT}$

#### 4.2.1 Tests

We tested our technique with different images processing attacks [19] :

- JPEG compression ;
- Adding noise (salt and pepper, gaussian) ;
- Median filtering ;
- Histogram equalization ;
- Geometrical attack (croping and rotation) ;
- Color reduction (GIF compression) ;

#### 4.2.2 Compression and Median Filtering

Robustness of the technique against :

- JPEG compression was tested by computing the correlation for different values of the Quality Factor (QF) between 10 % and 90 %.

- Median filtering was tested with filter mask from 3×3 to 9×9. Figures 3 and 4 below summarize the results.

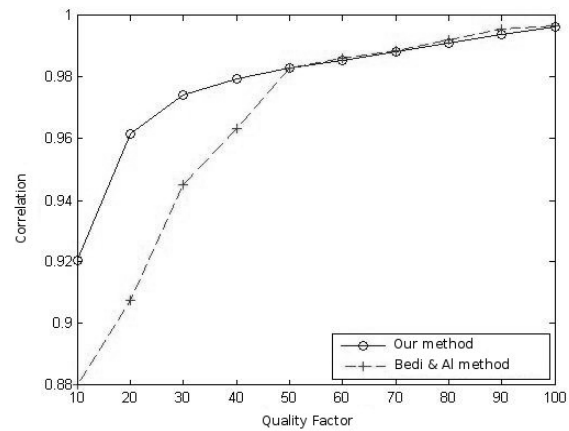

Fig. 3  Variation of the correlation based on the JPEG compression

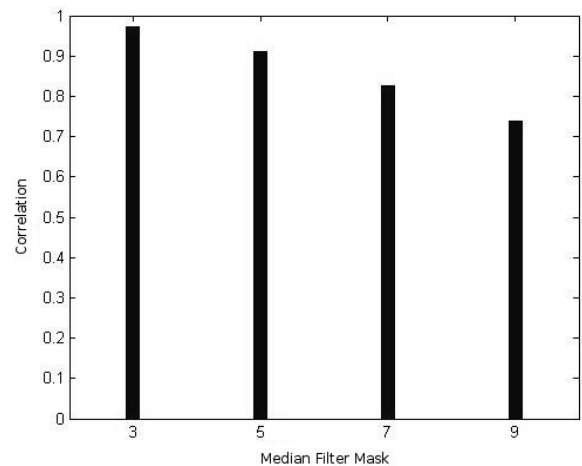

Fig. 4  Variation of the correlation depending on the size of the median filter

Figure 3 shows robustness against compression ; whatever the selected Quality Factor, the correlation value exceeds 0,9. In terms of filtering (Figure 4), the minimum value of correlation corresponding to a mask 9×9 is 0.7, then this shows robustness against filtering.

The following figures show the detector response on a database of 100 images for different types of attacks.

JPEG QF=10% (QF : Quality Factor)

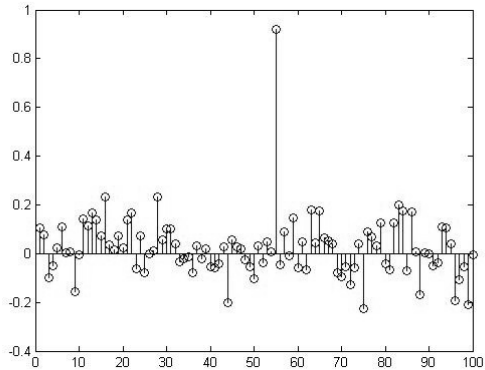

Gaussian noise V=0,03 (V : Variance)

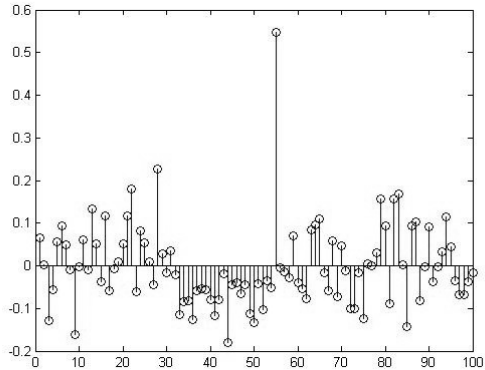

Salt & Pepper D=0,03 (D : Density)

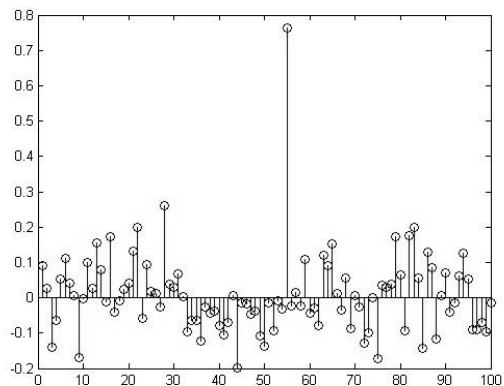

Median Filter 9x9

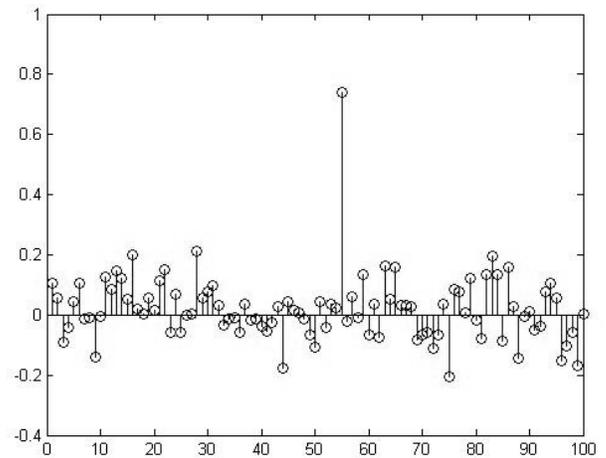

Cropping (remove 512x8 pixels)

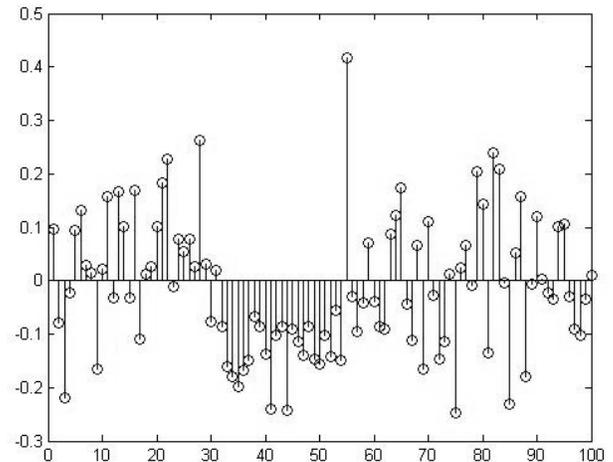

Rotation 1° to 2°

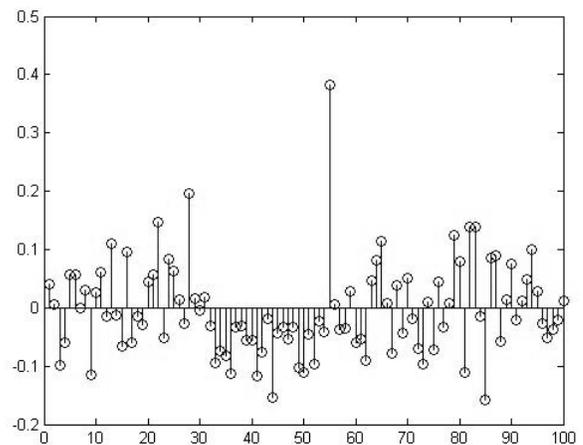

Histogram equalization

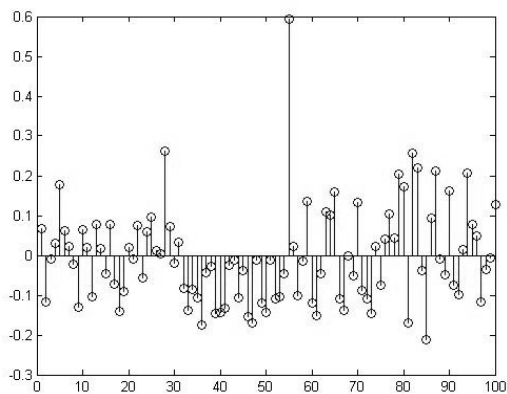

GIF colors reduction

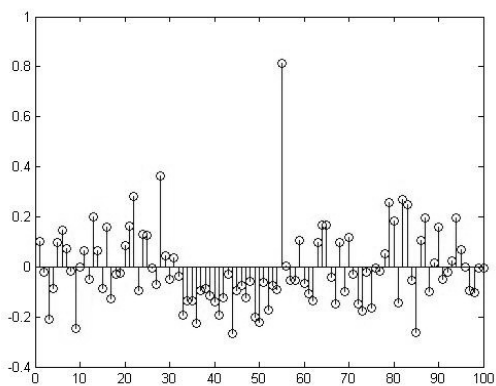

Fig. 5  Detector response for each type of attacks

For all types of attacks, we always note a "peak" on number 55 (mark detected). Regarding the geometric attacks : cropping, in which 8 columns and 8 lines in the image have been removed, gives a correlation value 0.4, low rotations (1° to 2°) also give a similar result . This value is low, but its peak that stands out from the other image shows the presence of the mark in the host image.
The following tables show, respectively, the test images and some samples attacked with obtained mark images.

Table 1: Test images

| 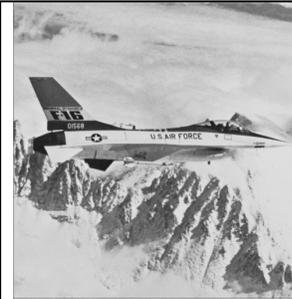 | 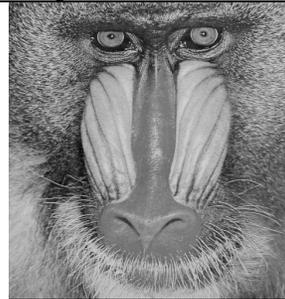 |
|---|---|
| Plane | Mandrill |

For each type of attack, we put in the second column of the extracted mark.

Table 2: Outline of the attacked images and the obtained mark

| Attacked image | Obtained mark |
|---|---|
| 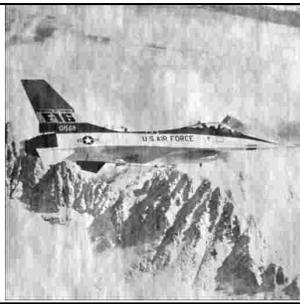 | 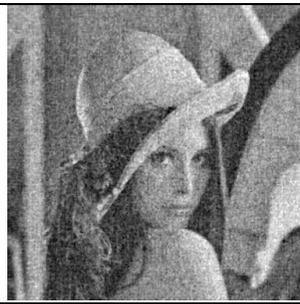 |
| JPEG QF=10% | |
| 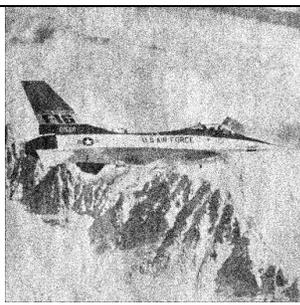 | 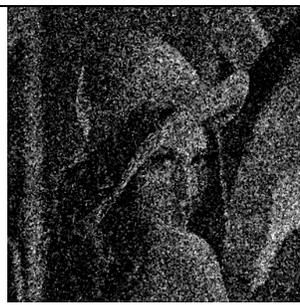 |
| Gaussian noise V=0,03 | |
| 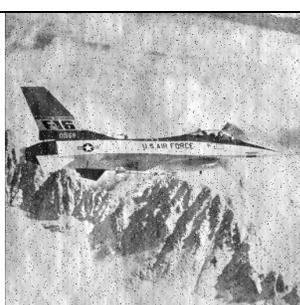 | 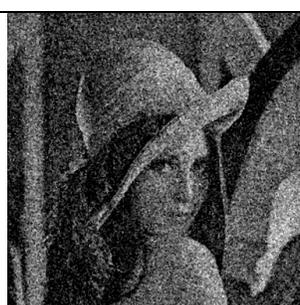 |
| Salt & Pepper noise D=0,03 | |
| 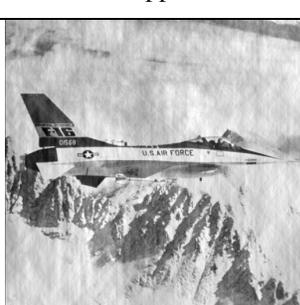 | 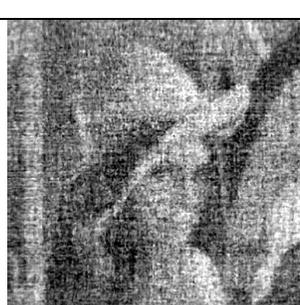 |
| Median Filtrage 9x9 | |

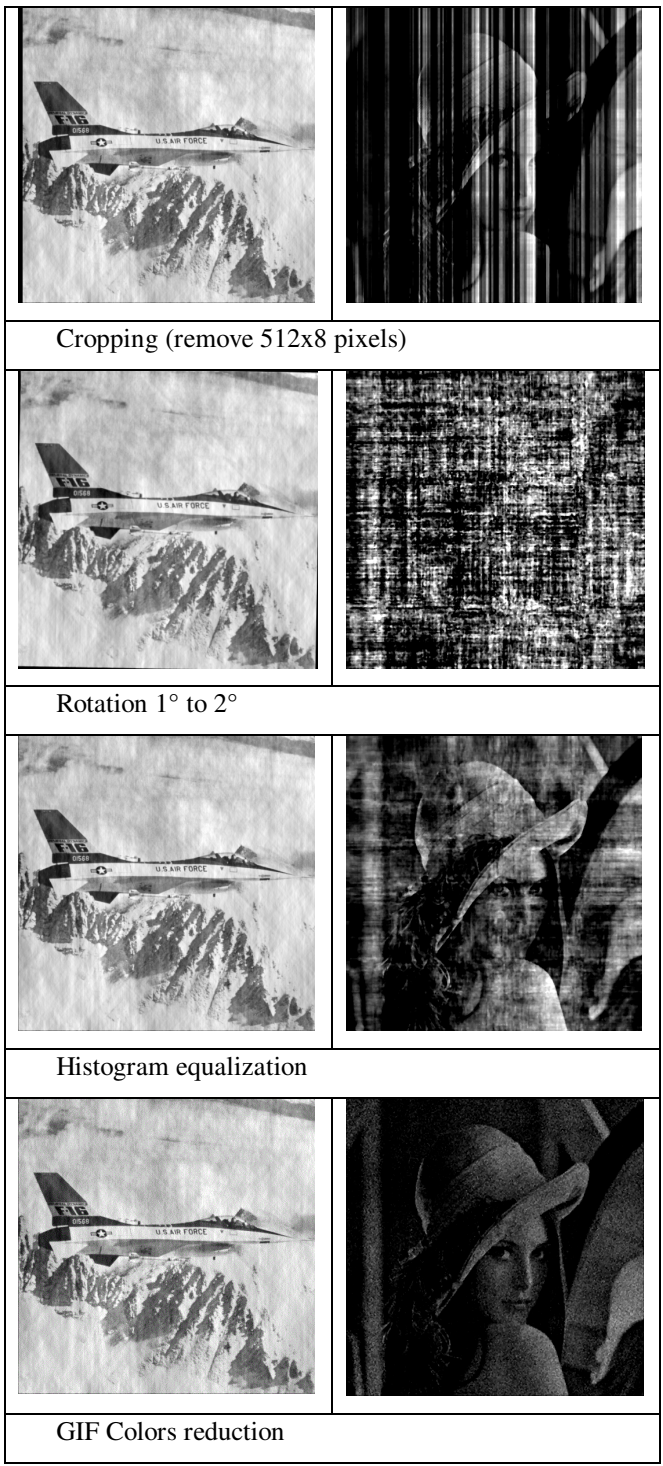

Cropping (remove 512x8 pixels)

Rotation 1° to 2°

Histogram equalization

GIF Colors reduction

### 4.3 Discussion

These results show that the method is robust against different types of attacks except geometric attacks (more than 2° rotations), which completely destroyed the brand. Compared to the Bedi et al [16], Jiansheng [13] and Saryazdi [7] methods, the obtained correlation coefficient after compression has been improved. With the Bedi's method, for a quality factor of 20%, the correlation is 0.9076, Saryazdi's method gives 0.8955 for a quality factor of 40%. Figure 3 shows the improvement in the quality factors less than 50% but for the rest, we find the same correlation factors with the Bedi's method.

Our technique greatly improves the watermark capacity because it allows inserting a grayscale image of the same size as the host image. Robustness could be improved by increasing the coefficient α but this operation will decrease the imperceptibility of the brand.

As Schur decomposition works only with square matrix, for processing rectangular images, simply break them down into several square blocks and insert the mark in each block.

Our method cannot solve all the problems of watermarking because it also has its limits. Oliveria [20] says "an algorithm that resists all types of attacks does not exist"

## 5. Conclusion and Perspectives

In this paper, we proposed a "robust" watermarking method that adds $T_W$ component of the mark, after Schur decomposition, in the DCT of the host image. $U_W$ component of the brand is used as a key for decoding because the watermarker must possess it before extraction. We could use our technique to watermark digital signatures with the photo of the owner, but we need to improve resistance to geometric attacks such as rotation.
Our outlook will turn to a variant of the method that is very robust against the geometric attacks by combining [21] the transformations that are resistant to geometric attacks such as Fourier transform. We could also consider replacing the DCT by the DLT transform [22] which is much more efficient. Finally to improve the imperceptibility, we could imagine a scheme based on WHT transform [23].

**Acknowledgments**

Authors thank IST-D (Institut Supérieur de Technologie d'Antsiranana) for its Sponsor and Financial Support.

**Henri Bruno Razafindradina** was born in Fianarantsoa, Madagascar, on 1978. He received, respectively, his M.S degree and Ph.D in Computer Science and Information Engineering in 2005 and 2008. He served since 2010 as a professor at Higher Institute of Technology Diego Suarez, became an assistant lecturer in 2011. His current research interests include images compression, multimedia, computer vision, information hiding.

**Nicolas Raft Razafindrakoto** is a professor at the Polytechnic High School of Antananarivo. His current research interests include petri networks and computer science.

**Paul Auguste Randriamitantsoa** is a professor at the Polytechnic High School of Antananarivo. His current research interests include automatic and computer science.